\documentclass[10pt]{article}
\usepackage[margin=0.85in]{geometry}
\usepackage{times}
\usepackage{microtype}
\usepackage{booktabs}
\usepackage{array}
\usepackage{tabularx}
\usepackage{enumitem}
\usepackage{xurl}
\usepackage[hidelinks]{hyperref}
\usepackage{natbib}
\usepackage{xcolor}
\usepackage{amsmath}
\usepackage{amssymb}
\usepackage{graphicx}
\usepackage{tikz}
\usetikzlibrary{arrows.meta,positioning,calc}
\usepackage{float}
\usepackage{caption}
\setlist{nosep,leftmargin=*}
\newcolumntype{Y}{>{\raggedright\arraybackslash}X}
\newcolumntype{P}[1]{>{\raggedright\arraybackslash}p{#1}}
\definecolor{navy}{RGB}{27,60,94}
\definecolor{steel}{RGB}{80,100,119}
\definecolor{lightblue}{RGB}{238,245,251}
\definecolor{lightgray}{RGB}{245,246,247}
\hypersetup{pdftitle={Artificial Intelligence in Finance and the Early-Career Skills Shift: A Structured Evidence Synthesis of Decision Delegation, Bank Deployments, and Business Education},pdfauthor={Prakhar Singh Sisodia}}

\title{Artificial Intelligence in Finance and the Early-Career Skills Shift:\\
A Structured Evidence Synthesis of Decision Delegation, Bank Deployments, and Business Education}
\author{Prakhar Singh Sisodia\\Boston University\\\texttt{prakhars@bu.edu}}
\date{October 2026}

\begin{document}
\maketitle

\begin{abstract}
Artificial intelligence is becoming a general interaction and decision-support layer across finance. Banks use AI for research retrieval, document synthesis, meeting follow-up, customer service, coding, fraud detection, risk support, and increasingly multi-step workflows. This creates a practical question for firms, regulators, universities, and early-career professionals: which parts of finance work can be delegated safely, which remain human-owned, and how should learning change when AI compresses the tasks through which novices traditionally build judgment?

This paper presents a structured evidence synthesis of research and public evidence released from 2023 through October 2026 across finance, labor economics, education, bank deployments, and financial regulation. Numerical results and deployment claims remain attributed to their original sources. The synthesis yields four connections. First, AI is strongest at retrieval, transformation, coding, and first-pass analysis, while ambiguous judgment, fiduciary responsibility, client trust, and exception ownership remain harder to delegate. Second, financial institutions are moving toward constrained autonomy rather than unrestricted automation. Third, productivity assistance can weaken apprenticeship if output improves faster than unaided capability. Fourth, decision delegation should depend jointly on consequence, reversibility, verification strength, and operational resilience.

The paper develops a five-layer Finance Work Stack, a Decision Delegation Matrix, the AI Apprenticeship Paradox, an Apprenticeship Preservation Protocol, and a Career Readiness Stack. It argues that the central challenge is not whether AI will replace finance, but how institutions can increase useful automation without weakening accountability, human capability, cyber resilience, or the future talent pipeline.
\end{abstract}


\section{Introduction}

Finance has used algorithms for decades. Credit scoring, fraud detection, quantitative trading, portfolio optimization, and risk modeling long predate the current wave of generative artificial intelligence. What changed after 2022 is the interface and the scope. Large language models (LLMs) can work directly with natural language, code, documents, tables, and increasingly tools. That makes AI relevant not only to quantitative specialists, but to the ordinary workflow of bankers, analysts, advisors, accountants, risk teams, and business students.

The change is already visible in institutional deployment. JPMorganChase's LLM Suite reached 200,000 onboarded users within eight months of its 2024 release \citep{jpm2025llmsuite}. Morgan Stanley deployed a research assistant for investment banking, sales and trading, and research, and separately rolled out advisor tools for internal knowledge retrieval and meeting follow-up \citep{morgan2023assistant,morgan2024debrief,morgan2024researchgpt}. Bank of America reports that more than 90\% of employees use its internal AI-driven virtual assistant and that its client-facing Erica assistant has passed three billion interactions \citep{bofa2025workforce,bofa2025erica}. Goldman Sachs describes developer copilots and a natural-language internal assistant in its annual report, followed by a firmwide rollout of the GS AI Assistant in 2025 \citep{goldman2024annual,goldman2025assistant}.

The adoption pattern is broader than a few U.S. banks. A 2024 Bank of England and Financial Conduct Authority survey found that 75\% of responding financial firms were already using AI and another 10\% planned to do so within three years. Foundation models accounted for 17\% of reported AI use cases. Importantly, 55\% of use cases had some degree of automated decision-making, but only 2\% were described as fully autonomous \citep{boe2024ai}. That distinction matters: finance is becoming more automated without becoming fully unsupervised.

For students, the labor-market signal is equally important. \citet{eisfeldt2025genai} report that demand for AI skills in finance and insurance job postings roughly tripled between ChatGPT's release and mid-2024, while finance occupations were almost twice as exposed to generative-AI productivity effects as the economy-wide average. The exposure is not uniform. Jobs with a larger interpersonal component, such as financial advising and some managerial roles, have less core-task exposure than occupations centered more heavily on codifiable analytical work.

This paper asks a practical question: \emph{what does the combined research imply for a business student entering finance?} The answer cannot come from one benchmark or one bank announcement. It requires connecting several literatures that are usually discussed separately: AI capability, finance-specific AI, enterprise deployments, labor-market exposure, human-AI productivity, education, and financial regulation.

\subsection*{Claim and attribution boundary}

This is a structured evidence synthesis and research agenda. It does not report original model benchmarks, bank productivity measurements, labor-market estimates, or educational experiments. All numerical results and descriptions of existing systems belong to the cited researchers and institutions. The original contribution is the organization of those results into the Finance Work Stack, Decision Delegation Matrix, AI Apprenticeship Paradox, Apprenticeship Preservation Protocol, Career Readiness Stack, and a set of testable questions for finance education and responsible delegation.

\section{Scope, Search Strategy, and Evidence Coding}

The review prioritizes material published or publicly released from 2023 through October 7, 2026. Sources were identified through targeted searches of scholarly databases and official publication sites using combinations of terms such as \emph{AI in finance}, \emph{financial services generative AI}, \emph{bank AI deployment}, \emph{finance future of work}, \emph{AI education}, \emph{agentic finance}, and \emph{financial AI governance}. Backward citation tracing was used for foundational studies, while official bank, regulator, and laboratory pages were used to verify deployment and policy claims.

Sources fall into four evidence groups:
\begin{enumerate}
    \item \textbf{Peer-reviewed and working-paper research} from leading universities and research groups, including Stanford, MIT, Harvard, Wharton/University of Pennsylvania, UCLA Anderson, and other finance, labor-economics, and education researchers.
    \item \textbf{Financial-system and regulatory research} from the Bank for International Settlements (BIS), Financial Stability Board (FSB), Bank of England, Financial Conduct Authority, and related institutions.
    \item \textbf{Official deployment evidence} from financial institutions such as JPMorganChase, Morgan Stanley, Bank of America, and Goldman Sachs. These sources establish what was deployed and how institutions describe its use; they are not treated as independent causal evidence of productivity.
    \item \textbf{Technical work and benchmarks} from organizations such as Bloomberg, Stanford CRFM, Microsoft Research, and Mastercard where they directly illuminate capability, evaluation, safety, work patterns, or fraud.
\end{enumerate}

Each included source was coded along six dimensions shown in Table~\ref{tab:coding}. This coding is designed to keep deployment scale, causal evidence, and forward-looking interpretation separate.

\begin{table}[H]
\centering
\caption{Evidence coding used in the structured synthesis.}
\label{tab:coding}
\small
\begin{tabularx}{\textwidth}{P{0.20\textwidth} Y}
\toprule
Dimension & Coding question \\
\midrule
Evidence type & Is the source peer-reviewed research, a working paper, regulatory evidence, an official deployment report, or a market/education survey? \\
Finance work layer & Does the evidence concern retrieval, transformation, analysis, recommendation, or accountable decision ownership? \\
Delegation level & Is AI assisting, recommending, executing within boundaries, or acting with material discretion? \\
Human responsibility & Which tasks remain human-owned: verification, sign-off, client trust, exception management, or governance? \\
Failure and risk mode & What can go wrong: hallucination, numerical inconsistency, bias, privacy loss, cyber risk, correlated behavior, or skill atrophy? \\
Transferability & Is the result causal and broadly tested, context-specific, organization-reported, or a forecast requiring caution? \\
\bottomrule
\end{tabularx}
\end{table}

The synthesis is structured rather than PRISMA-complete and does not claim exhaustive coverage. Marketing claims are separated from controlled studies. Where evidence has been challenged or withdrawn, it is excluded from the empirical basis. For example, a widely discussed University of Chicago working paper on LLM financial-statement analysis was withdrawn after a co-author reported replication inconsistencies; its performance claims are therefore not used here \citep{boothwithdrawal}.

\section{From Traditional Financial AI to the Generative-AI Layer}

\subsection{Finance was already an AI-heavy industry}

The current wave should not be confused with the first use of AI in finance. Traditional machine learning has long supported fraud detection, underwriting, trading, risk, and customer segmentation. BIS researchers describe the financial system as an information-processing system and trace a progression from rule-based systems to machine learning, generative AI, and emerging agents \citep{aldasoro2024intelligent}.

The practical difference is that generative AI can operate on the unstructured information that fills much of financial work: filings, research reports, policies, emails, meeting transcripts, client notes, news, legal documents, and code. That moves AI closer to the daily work of analysts and relationship managers.

\subsection{Domain-specific language models and financial evaluation}

BloombergGPT was an early example of a finance-specific LLM trained on a mixture of large financial and general-purpose corpora. The paper showed that domain data could improve financial-language performance without completely sacrificing general capability \citep{wu2023bloomberggpt}. The broader lesson is not that every bank needs to train a 50-billion-parameter model. It is that finance benefits from domain context, terminology, and evaluation.

Evaluation is becoming an infrastructure problem of its own. Stanford's HELM Finance benchmark, produced in collaboration with Wells Fargo, evaluates leading models on finance-oriented tasks using real financial documents and transparent prompting \citep{stanfordhelmfinance}. This is important because a model that is fluent in ordinary conversation may still fail on financial nuance, numerical consistency, source grounding, or policy constraints.

\subsection{The rise of retrieval and internal knowledge assistants}

The first large-scale generative-AI deployments in finance are often less about autonomous investment decisions than about finding and transforming information. Morgan Stanley's early GPT-4 initiative was explicitly designed to let financial advisors query the firm's own intellectual capital and receive answers linked back to source documents \citep{morgan2023assistant}. Its later AskResearchGPT product extended this pattern to investment banking, sales and trading, and research, drawing from tens of thousands of proprietary reports \citep{morgan2024researchgpt}.

JPMorganChase's LLM Suite similarly provides a controlled internal environment for idea generation, drafting, and access to language models, with future plans to integrate more internal data and multi-step agents \citep{jpm2025llmsuite}. This pattern is best understood as \emph{institutional retrieval plus synthesis}: the model becomes an interface to a bank's private knowledge base.

There is also a safety lesson. Bloomberg researchers reported that retrieval-augmented generation (RAG) does not automatically make an LLM safer and introduced finance-specific risk thinking for generative systems \citep{bloomberg2025safety}. In finance, adding more context can improve factual grounding while simultaneously creating new pathways for unsafe or inappropriate outputs. Retrieval therefore needs permissions, provenance, policy filters, and testing, not just vector search.

\section{Where AI Is Being Used Across Finance}

The deployment landscape can be organized into seven recurring categories.

\subsection{Research, search, and synthesis}

LLMs are increasingly used to search internal research, summarize filings and reports, compare documents, extract key facts, and draft first-pass analyses. These uses directly target one of the most time-consuming parts of junior knowledge work: locating relevant information and turning it into a usable format. Morgan Stanley's AskResearchGPT and the information-retrieval portion of its advisor assistant are direct examples \citep{morgan2023assistant,morgan2024researchgpt}.

\subsection{Writing, presentations, and document workflows}

Financial work produces a large volume of text: client emails, investment memos, pitch materials, committee papers, policy notes, call summaries, and internal updates. Goldman Sachs' internal assistant is described as supporting document summarization, drafting, data analysis, and related knowledge-work tasks \citep{goldman2025assistant}. Wharton's enterprise-adoption research similarly identifies writing/editing, data analysis, and summarization among common generative-AI use cases \citep{wharton2025adoption}.

The key shift is from blank-page creation to human review. A junior employee may spend less time writing the first version and more time checking facts, adjusting tone, reconciling numbers, and making the output defensible.

\subsection{Meeting intelligence and client follow-up}

Morgan Stanley's Debrief shows how generative AI can convert a live client interaction into structured notes, action items, an editable follow-up email, and a CRM record \citep{morgan2024debrief}. This is a concrete example of workflow compression: several administrative tasks that traditionally happen after a meeting are merged into one assisted pipeline.

For wealth management, the implication is not simply fewer notes. It shifts more of the advisor's scarce attention toward the conversation itself. Morgan Stanley explicitly frames these systems as augmenting the advisor while preserving the human relationship \citep{morgan2023assistant}.

\subsection{Customer service and personalization}

Bank of America's Erica demonstrates a different scale of AI adoption: consumer-facing conversational support integrated with ordinary banking. The bank reports more than three billion client interactions by 2025, along with deployment across Merrill, private banking, employee support, and corporate banking contexts \citep{bofa2025erica}. The internal version is used by more than 90\% of employees according to Bank of America, with the bank reporting a large reduction in IT-service-desk calls \citep{bofa2025workforce}.

The important career implication is that customer service is not disappearing; its composition changes. Routine questions become cheaper to answer automatically, while exceptions, complaints, complex products, and emotionally sensitive decisions move toward humans.

\subsection{Fraud, AML, cybersecurity, and risk}

Traditional AI remains especially strong where institutions have large labeled datasets and repetitive detection problems. The Bank of England/FCA survey identified data and analytical insights, anti-money-laundering/fraud, and cybersecurity among the highest perceived current benefits of AI \citep{boe2024ai}. Mastercard's 2025 fraud research similarly describes the use of AI to detect subtle transaction patterns while emphasizing governance, shared intelligence, and false-positive management \citep{mastercard2025fraud}.

These systems do not eliminate risk professionals. They change the unit of work from manually reviewing every item to designing thresholds, investigating exceptions, monitoring model drift, and defending decisions to regulators and customers.

\subsection{Credit, lending, and underwriting}

AI can process hard, codifiable information efficiently, but finance also depends on soft information. BIS research on relationship lending studies how AI-based credit scoring interacts with traditional relationship-based information rather than treating one as a complete replacement for the other \citep{gambacorta2025relationship}. This is a useful general pattern: the more a decision depends on context, trust, or information that is difficult to encode, the stronger the case for a hybrid workflow.

\subsection{Trading and agentic financial decisions}

The frontier becomes more consequential when AI is allowed to act. Wharton researchers studying reinforcement-learning traders show that AI agents can create market-efficiency and collusion concerns even without explicit communication or intent \citep{dou2025trading}. BIS work on AI agents for cash management in payment systems illustrates a more operational version of the same shift: models are moving from producing information to making sequential decisions in constrained financial environments \citep{bis2025agents}.

This is where governance becomes central. A system that drafts a memo can be checked before use. A system that allocates liquidity or submits an order changes the state of the world. The controls therefore need to become stricter as AI moves from advice to action.

\section{The Finance Work Stack}

The research and deployments suggest a useful way to decompose finance work. Rather than asking whether an occupation will be automated, it is more precise to ask which layer of a workflow AI can perform reliably.

\begin{figure}[ht]
\centering
\fbox{\begin{minipage}{0.92\linewidth}
\centering
\textbf{Layer 5: Own the Decision}\\
Accountability, fiduciary duty, sign-off, client trust, exception handling\\[4pt]
$\uparrow$\\[-2pt]
\textbf{Layer 4: Recommend and Exercise Judgment}\\
Scenario choice, trade-offs, negotiation, risk appetite, narrative conviction\\[4pt]
$\uparrow$\\[-2pt]
\textbf{Layer 3: Analyze}\\
Valuation, forecasting, sensitivity analysis, classification, anomaly detection\\[4pt]
$\uparrow$\\[-2pt]
\textbf{Layer 2: Transform}\\
Summarize, compare, draft, extract, code, populate templates, reconcile formats\\[4pt]
$\uparrow$\\[-2pt]
\textbf{Layer 1: Retrieve}\\
Search research, policies, filings, market data, prior work, client history
\end{minipage}}
\caption{The proposed Finance Work Stack. AI capability is advancing upward, but the evidence from current bank deployments and regulatory surveys suggests that full delegation remains rare at the top layers.}
\label{fig:stack}
\end{figure}

\textbf{Layer 1 - Retrieve.} AI is already strong at finding relevant material when the information environment is accessible and permissions are well designed.

\textbf{Layer 2 - Transform.} Summarization, drafting, document comparison, coding, translation, note generation, and formatting are among the most mature generative-AI use cases.

\textbf{Layer 3 - Analyze.} Capability is growing quickly, but this layer demands stronger numerical checks, data lineage, and evaluation. Finance-specific benchmarks exist precisely because fluent text is not enough \citep{stanfordhelmfinance}.

\textbf{Layer 4 - Recommend.} Recommendation combines analysis with objectives, constraints, and uncertainty. The model can produce options, but the quality of the decision increasingly depends on framing the right problem and recognizing where the model is outside its competence.

\textbf{Layer 5 - Own the decision.} Accountability is the hardest layer to automate. The Bank of England/FCA finding that only 2\% of surveyed AI use cases were fully autonomous is consistent with the view that institutions are automating pieces of the stack faster than they are delegating final responsibility \citep{boe2024ai}.

This stack also explains why job titles can survive even when workflows change dramatically. A financial analyst can remain a financial analyst while spending far less time on retrieval and formatting and more time on assumptions, validation, and communication.

This stack describes \emph{what} work is being performed. A second framework is needed to decide \emph{how much authority} should be delegated. Figure~\ref{fig:delegation} combines two variables that recur across the evidence: the consequence and reversibility of a decision, and the strength of independent verification.

\begin{figure}[H]
\centering
\begin{minipage}{0.96\linewidth}
\centering
\small
\begin{tabularx}{\linewidth}{|P{0.16\linewidth}|Y|Y|}
\hline
 & \textbf{Weak independent verification} & \textbf{Strong independent verification} \\
\hline
\textbf{Higher consequence / lower reversibility} &
\textbf{Human-owned.} AI is limited mainly to retrieval, transformation, and scenario support; people retain decision authority. &
\textbf{Co-decision.} AI may recommend or prepare an action, but human sign-off, provenance, and an audit trail remain mandatory. \\
\hline
\textbf{Lower consequence / higher reversibility} &
\textbf{Assist / experiment.} AI output is checked before use while evidence and controls are still developing. &
\textbf{Bounded execution.} Deterministic checks, limits, monitoring, and rollback allow selected actions to execute automatically. \\
\hline
\end{tabularx}
\end{minipage}
\caption{Decision Delegation Matrix. This is a synthesis framework, not a regulatory standard. Delegation should increase only when the decision is sufficiently verifiable, reversible, and bounded relative to its consequence.}
\label{fig:delegation}
\end{figure}

The matrix explains why document summarization and meeting follow-up can be automated earlier than credit approval, trading, suitability, or liquidity decisions. It also avoids a false binary between ``human'' and ``autonomous.'' A high-consequence decision may use substantial AI analysis while preserving human ownership, whereas a low-consequence, highly verifiable action may be executed automatically within strict limits.

\section{What the Labor Research Implies for Early-Career Finance}

\subsection{Finance is highly exposed, but exposure is not the same as job loss}

A useful starting point is the distinction between task exposure and employment outcomes. \citet{eloundou2023gpts} estimate that LLMs could affect a meaningful share of tasks across a large part of the U.S. workforce, especially when combined with software. The paper explicitly does not predict an adoption timeline or direct job losses.

Finance-specific evidence sharpens the picture. \citet{eisfeldt2025genai} find that finance occupations are highly exposed to generative-AI productivity effects and that finance-sector demand for AI skills rose rapidly after ChatGPT. They also find meaningful variation within finance. Core tasks in occupations with more nonroutine interpersonal work are less exposed than tasks centered on codifiable analysis. This helps explain why relationship-heavy roles can remain valuable even as analytical support becomes cheaper.

\subsection{The early-career signal is more concerning}

Stanford researchers using high-frequency payroll data report that workers aged 22--25 in the most AI-exposed occupations experienced a 13\% relative decline in employment after widespread generative-AI adoption, while older workers in the same occupations and workers in less-exposed occupations were more stable \citep{brynjolfsson2025canaries}. This result is not finance-specific and should not be presented as a forecast for investment banking or accounting. It is, however, relevant because many entry-level finance roles are built around exactly the kinds of information-processing tasks that LLMs can accelerate.

The mechanism may be simple: firms can obtain more output from smaller junior teams when drafting, search, basic analysis, and coding become faster. But that is only one possible response. Firms may also keep team size constant and increase the quantity or depth of work. The net employment effect remains an empirical question.

\subsection{AI can disproportionately help novices}

The other side of the story is augmentation. In a large customer-support field study, \citet{brynjolfsson2025work} find average productivity gains from a generative-AI assistant, with much larger gains for novice and lower-skilled workers than for experienced workers. The authors interpret the tool partly as a mechanism for diffusing the practices of stronger workers.

This is highly relevant to finance apprenticeships. An entry-level analyst with a well-designed internal assistant may gain immediate access to examples, explanations, prior analyses, code snippets, and institutional knowledge that previously took months to learn informally. AI can flatten the early learning curve.

\subsection{But the technology frontier is jagged}

The HBS/Wharton/MIT/BCG field experiment led by \citet{dellacqua2023jagged} provides an important warning. On tasks inside the model's capability frontier, consultants using GPT-4 completed more work, worked faster, and produced higher-quality outputs. On a task deliberately selected outside that frontier, AI users were less likely to reach the correct answer. The lesson is not merely ``AI makes people faster.'' It is that users must know when the tool is likely to fail.

This creates a new professional skill: \textbf{frontier recognition}. The valuable analyst is increasingly the person who knows which parts of a task can be delegated, which require independent reconstruction, and which need an expert review.

\section{The AI Apprenticeship Paradox}

The combined labor and education evidence points to a tension that is especially important for business schools.

\begin{quote}
\textbf{AI apprenticeship paradox:} The workers who can benefit most from AI assistance are often the same workers who traditionally learn by performing the tasks that AI is beginning to automate.
\end{quote}

There is evidence for both sides. Generative AI can help novices perform closer to expert practice \citep{brynjolfsson2025work}. At the same time, a controlled education experiment by \citet{bastani2025learning} found that unrestricted GPT access improved practice performance but could reduce later unaided performance; a tutor design with learning safeguards mitigated much of that harm. The setting was high-school mathematics rather than finance, so the result should not be mechanically generalized to business education. The mechanism, however, is directly relevant: if students outsource the cognitive struggle required to form a skill, short-run output can improve while long-run capability weakens.

University usage is already widespread. Stanford's 2026 AI Index reports that roughly four in five U.S. high-school and college students use AI for school-related tasks \citep{stanford2026index}. A Harvard undergraduate survey similarly found very high adoption and evidence that some students substituted AI for office hours or assigned reading \citep{hirabayashi2024harvard}. At HBS, a course-specific tutor used by 930 introductory accounting students generated more than 20,000 prompts in one term, showing how quickly course-grounded assistants can become part of the learning environment \citep{hbs2025tutor}.

For finance education, the wrong response is either total prohibition or unrestricted outsourcing. A better model is staged use:

\begin{enumerate}
    \item \textbf{Build the primitive manually.} Students should be able to read the three financial statements, build a basic valuation, calculate returns, reason about risk, and explain accounting links without an AI completing the reasoning for them.
    \item \textbf{Use AI to accelerate execution.} Once the primitive is understood, AI can help write code, check formulas, summarize filings, generate scenario tables, and surface alternative explanations.
    \item \textbf{Require verification.} Students should cite sources, reconcile numbers, reproduce calculations, document prompts/tool use where appropriate, and explain why the final decision is defensible.
    \item \textbf{Practice unaided checkpoints.} Courses should preserve exams, oral defenses, whiteboard exercises, or timed cases where students demonstrate that the underlying skill has actually been acquired.
\end{enumerate}

This is consistent with the direction described by Chicago Booth faculty integrating AI into finance teaching while emphasizing both use and skepticism \citep{booth2025education}.

\subsection*{An Apprenticeship Preservation Protocol}
The paradox suggests that responsible AI education should measure learning transfer rather than assisted output alone. A practical protocol has five stages:
\begin{enumerate}
    \item \textbf{Unaided baseline:} complete the core accounting, valuation, risk, or market task without AI.
    \item \textbf{AI-assisted execution:} use approved tools to accelerate retrieval, coding, drafting, or scenario generation.
    \item \textbf{Adversarial verification:} inspect an output containing seeded numerical, citation, or interpretation errors.
    \item \textbf{Independent defense:} explain assumptions and reproduce key reasoning orally, on a whiteboard, or under timed conditions.
    \item \textbf{Transfer test:} apply the concept to a new company, product, or market setting after a delay.
\end{enumerate}
This protocol is a proposed educational design, not a claim of validated superiority. Its purpose is to preserve the cognitive practice needed for future oversight while still teaching modern workflows.

\section{A Career Readiness Stack for Business Students}

The evidence does not support the idea that every finance student should become an AI engineer. It does support a broader conclusion: the baseline for a competitive finance graduate is moving from ``finance plus spreadsheets'' toward ``finance plus data plus AI supervision.''

\subsection{Layer 1: Finance fundamentals}

Accounting, corporate finance, valuation, statistics, economics, market structure, portfolio theory, and risk remain foundational. AI increases the value of these basics because verification is impossible without a mental model of what a plausible answer looks like.

\subsection{Layer 2: Data and computational fluency}

Students do not need to be software engineers, but they should be comfortable with Excel, SQL, basic Python, APIs, data cleaning, and visualization. These skills make AI outputs more testable and allow the student to move from chat-only use toward reproducible workflows.

\subsection{Layer 3: AI workflow fluency}

This includes structured prompting, retrieval-augmented workflows, document extraction, tool use, agents, model comparison, and knowing when to use deterministic code rather than an LLM. The goal is not prompt tricks; it is reliable task decomposition.

\subsection{Layer 4: Verification, governance, and model risk}

Financial institutions operate under constraints that ordinary consumer AI use does not. BIS and FSI research emphasize data governance, model risk, privacy, third-party dependence, expertise, and oversight as central issues for financial AI \citep{crisanto2024regulating,aldasoro2024intelligent}. A job candidate who can explain hallucination risk, source provenance, permissions, auditability, reproducibility, and human escalation has a useful advantage.

\subsection{Layer 5: Human judgment and relationships}

As codifiable work gets cheaper, the relative value of ambiguity management, persuasion, negotiation, stakeholder alignment, empathy, and accountability rises. This is consistent with the occupation-level patterns in \citet{eisfeldt2025genai}: finance roles relying more heavily on nonroutine interpersonal skills tend to have lower core-task AI exposure.

\begin{table}[ht]
\centering
\caption{How the skill shift may appear across common early-career finance paths. ``AI now'' summarizes documented or strongly supported use patterns; it is not a claim that every firm has automated the task.}
\small
\begin{tabularx}{\textwidth}{P{2.2cm} Y Y Y}
\toprule
\textbf{Path} & \textbf{AI now / emerging} & \textbf{Human edge} & \textbf{Student signals worth building} \\
\midrule
Investment banking & Research retrieval, first drafts, document comparison, pitch support, data extraction & Deal judgment, client narrative, negotiation, assumption ownership & Accounting, valuation, Excel, Python/SQL basics, source verification, concise writing \\
\addlinespace
Equity / credit research & Search, summarization, screening, transcript/filing analysis, coding assistance & Causal thesis, variant perception, risk framing, conviction under uncertainty & Financial modeling, markets, statistics, Python, research design, evidence discipline \\
\addlinespace
Corporate finance / FP\&A & Variance explanation, reporting drafts, scenario generation, forecasting support & Business context, capital allocation, cross-functional influence & Accounting, budgeting, forecasting, BI tools, SQL, presentation skills \\
\addlinespace
Wealth management & Research Q\&A, meeting notes, follow-up drafts, personalization & Trust, behavioral coaching, suitability, life context, relationship management & Investments, tax/retirement basics, communication, CRM/data fluency, AI supervision \\
\addlinespace
Risk / credit & Document analysis, anomaly detection, scoring support, monitoring & Exceptions, fairness, policy interpretation, risk appetite, sign-off & Credit analysis, statistics, regulation, model risk, data lineage, Python/SQL \\
\addlinespace
Operations / compliance & Policy search, case triage, AML/fraud support, documentation & Investigation, escalation, control design, regulatory accountability & Process design, controls, analytics, governance, clear documentation \\
\addlinespace
Fintech / finance product & Natural-language interfaces, agent workflows, automated research and support & Product judgment, compliance-by-design, customer needs, system trade-offs & Finance domain depth, APIs, data, experimentation, AI evaluation, product thinking \\
\bottomrule
\end{tabularx}
\label{tab:roles}
\end{table}

\section{What Current Bank Deployments Reveal About the Future}

The most useful signal from bank deployments is not the name of any one product. It is the architecture of adoption.

\subsection{Pattern 1: AI is becoming an interaction layer}

Morgan Stanley explicitly describes a future in which AI sits between employees and applications such as CRM, research, reporting, risk analysis, and order-entry systems \citep{morgan2024debrief}. JPMorganChase describes LLM Suite as evolving toward an AI hub with more internal data and agents \citep{jpm2025llmsuite}. This suggests that employees may increasingly interact with enterprise systems through natural language rather than navigating each system directly.

For students, this means software fluency is changing. Knowing the menu path in a specific tool matters less than understanding the underlying data, process, and control.

\subsection{Pattern 2: The first target is workflow friction, not fiduciary judgment}

The strongest documented deployments focus on search, drafting, notes, support, and coding. This is rational. These tasks are high-volume, measurable, and usually reversible. They also sit lower in the Finance Work Stack.

As institutions gain confidence, systems will move upward. The Bank of England/FCA data already show some degree of automated decision-making in more than half of reported AI use cases, although complete autonomy remains rare \citep{boe2024ai}.

\subsection{Pattern 3: Internal data is the competitive asset}

General models are increasingly available to everyone. Banks differentiate by connecting them to proprietary research, policies, workflows, customer context, and historical decisions. Morgan Stanley's research corpus, Bank of America's client and service infrastructure, and JPMorganChase's planned internal-data integrations all illustrate this pattern \citep{morgan2024researchgpt,bofa2025erica,jpm2025llmsuite}.

The early-career implication is subtle: memorizing public information becomes less differentiating, while understanding proprietary context, data quality, and institutional process becomes more important.

\subsection{Pattern 4: Governance becomes part of ordinary finance work}

AI risk is not confined to a model-risk department. If employees use AI to draft a client communication, analyze a deal, or interpret policy, they participate in the control environment. Regulators emphasize human expertise, governance, data quality, explainability, third-party dependencies, and model risk \citep{crisanto2024regulating,boe2024ai}. These are becoming mainstream business skills.

\subsection{Pattern 5: Governance is becoming a lifecycle and resilience discipline}
Recent regulatory work strengthens the constrained-autonomy interpretation. The FSB's 2026 consultation proposes 12 sound practices spanning organization-wide governance, AI development and deployment, cyber and information-technology risk, and third-party dependencies \citep{fsb2026sound}. The report explicitly asks whether the practices are adequate for generative and agentic AI. In August 2026, the FSB Chair separately warned that frontier AI could alter the speed, scale, and economics of cyber risk and emphasized safe model release, deployment, and operational resilience \citep{fsb2026frontier}.

The implication is that responsible financial AI cannot be reduced to model accuracy. Institutions need lifecycle ownership, access control, provenance, incident response, fallback procedures, vendor concentration management, and recovery testing. As systems move from drafting to action, operational resilience becomes part of decision quality.

\section{A More Precise View of ``Automation''}

The phrase ``AI will automate finance'' hides several distinct mechanisms:

\begin{enumerate}
    \item \textbf{Time compression:} the same employee completes a task faster.
    \item \textbf{Quality augmentation:} the employee completes a task better because AI supplies examples, knowledge, or checks.
    \item \textbf{Task elimination:} a step disappears from the human workflow entirely.
    \item \textbf{Role rebundling:} the job remains, but its mix of tasks changes.
    \item \textbf{Headcount substitution:} the firm needs fewer people for the same output.
    \item \textbf{Demand expansion:} lower task cost leads the firm to do more analysis, serve more clients, or create new products.
\end{enumerate}

The available research strongly supports the first two mechanisms and provides emerging evidence for the third and fifth in some occupations, but it is too early to assume a single economy-wide outcome. Microsoft Research, for example, found that workers given a generative-AI tool changed independently adjustable behaviors such as email time more than coordination-heavy behaviors such as meetings \citep{dillon2025shifting}. That finding is consistent with the idea that automation arrives first where one worker can change the task without redesigning the organization.

For finance, this means the near-term effects may appear unevenly. A bank can deploy an email or document assistant quickly. Rebuilding a credit approval process, investment committee, or regulated customer journey around an AI agent requires new controls, accountability, and coordination.

\section{Research Questions for Finance Education and Early-Career Work}

The synthesis yields several questions that can be tested rather than debated abstractly.

\textbf{RQ1: Which entry-level finance tasks are being removed versus merely accelerated?}\\
A useful study would track task-level time allocation before and after enterprise AI deployment across analyst cohorts, separating retrieval, modeling, drafting, review, client interaction, and administrative work.

\textbf{RQ2: Does AI reduce or increase the speed at which junior finance professionals acquire independent judgment?}\\
The apprenticeship paradox predicts two competing effects: AI can expose novices to expert patterns faster, but excessive delegation can reduce productive struggle. Longitudinal studies should test unaided performance, not only assisted output.

\textbf{RQ3: Is there an AI-literacy wage or hiring premium within finance?}\\
\citet{eisfeldt2025genai} document rapid growth in AI-related skill demand. Future work can test whether AI skill mentions predict interview rates, starting compensation, faster promotion, or role mobility after controlling for finance knowledge and quantitative skills.

\textbf{RQ4: Which human skills become more valuable as analytical tasks become cheaper?}\\
The occupation-level evidence suggests a role for nonroutine interpersonal skills. Within finance, this can be tested across client-facing versus back-office roles, and across seniority.

\textbf{RQ5: How should universities design guarded AI use?}\\
Experiments can compare no-AI, unrestricted-AI, and structured-AI cohorts in accounting, valuation, statistics, and case analysis, measuring both immediate assignment quality and later unaided retention.

\textbf{RQ6: When does an AI agent cross the boundary from productivity tool to regulated decision-maker?}\\
The answer may depend less on model architecture than on whether the system changes customer outcomes, moves money, sets terms, or exercises discretion. This question connects technology design directly to governance.

\textbf{RQ7: Do finance firms respond to AI by shrinking junior cohorts or by raising the complexity of junior work?}\\
The two strategies have very different implications for talent pipelines. A smaller analyst class may save cost today but weaken the future supply of experienced professionals unless firms create alternative apprenticeship mechanisms.

\textbf{RQ8: Can decision-delegation readiness be measured consistently across financial workflows?}\\
Future work should operationalize consequence, reversibility, verification strength, auditability, cyber exposure, and human escalation, then test whether those measures predict safe deployment outcomes.

\textbf{RQ9: How should AI productivity be adjusted for operational and cyber resilience?}\\
A workflow that saves analyst time but raises outage, third-party, or incident-recovery risk may not create net institutional value. Studies should measure resilience-adjusted productivity rather than task speed alone.

\section{Practical Curriculum Implications}

A business-school response should be concrete. Based on the evidence reviewed here, an AI-ready finance curriculum would include five types of work.

\textbf{1. Manual foundations.} Students should still build models from blank spreadsheets, reconcile statements, perform time-value-of-money calculations, derive valuation logic, and explain core market mechanisms without AI assistance.

\textbf{2. AI-assisted analysis.} Students should use LLMs to summarize filings, draft research questions, generate code, compare scenarios, and critique their own work. The objective should be speed plus breadth, not substitution for reasoning.

\textbf{3. Verification assignments.} A useful finance-specific exercise is to give students an AI-generated memo containing subtle numerical, citation, and interpretation errors and grade them on detection and correction.

\textbf{4. Workflow projects.} Students should build small end-to-end systems: for example, ingest a public filing, retrieve relevant sections, extract metrics with deterministic code, ask an LLM for a narrative, verify the numbers, and produce a cited investment or credit memo.

\textbf{5. Governance and communication.} Students should learn to document AI use, explain model limitations, protect confidential data, design human escalation, and defend a decision orally. These are not peripheral ethics topics; they are operating requirements in regulated finance.

A curriculum like this does not require turning a finance degree into computer science. It makes finance students capable of supervising the systems that increasingly surround their work.

\section{Risks and Limits of an AI-Heavy Finance Future}

\subsection{Hallucination and numerical reliability}

LLMs can produce plausible but incorrect statements. In finance, errors can propagate into valuations, customer communications, compliance decisions, or trades. This makes source linkage and deterministic calculation especially important.

\subsection{Automation bias}

The jagged-frontier evidence shows that people can become less accurate when they trust AI on tasks outside its capability \citep{dellacqua2023jagged}. The more polished the output, the easier it can be to accept without reconstructing the logic.

\subsection{Data and privacy risk}

Banks work with confidential customer, market, and transaction data. Third-party model concentration and incomplete understanding of external systems are explicit concerns in the Bank of England/FCA survey \citep{boe2024ai}. Students entering finance need to understand that copying sensitive information into an unapproved chatbot is not simply bad practice; it can create legal and security exposure. The FSB has also identified frontier-AI cyber capability as an immediate concern for the financial system, making model access, release, recovery, and third-party resilience first-order governance questions \citep{fsb2026frontier}.

\subsection{Market and systemic risk}

If many institutions deploy similar AI models or agents, behavior can become correlated. The Wharton trading-collusion work shows one way algorithmic interaction can produce undesirable market outcomes without an explicit human conspiracy \citep{dou2025trading}. BIS research likewise treats AI as a potential source of both efficiency and new financial-stability concerns \citep{aldasoro2024intelligent}.

\subsection{Skill atrophy}

Education research suggests that assisted performance should not be mistaken for learned capability \citep{bastani2025learning}. In a professional setting, this risk matters because humans are expected to take over when systems fail. A workforce that has never practiced the underlying task may be less capable exactly when supervision is most needed.

\subsection{Unequal access to high-quality AI}

Large banks can build private retrieval systems, governance layers, proprietary datasets, and custom evaluations that students and smaller firms cannot easily reproduce. This could create a new form of institutional advantage. Universities can partially offset it by teaching transferable concepts rather than vendor-specific interfaces.

\section{Limitations and Responsible Claims}

Several limitations matter.

First, the paper combines different evidence types. Randomized field experiments, occupational exposure measures, regulatory surveys, and bank product announcements should not be treated as equally causal. Bank deployment statistics show scale of use, not necessarily productivity or employment effects.

Second, many labor studies are economy-wide rather than finance-specific. The Stanford early-career employment result is therefore used as a warning signal, not as proof that young finance employment has declined by the same amount \citep{brynjolfsson2025canaries}.

Third, AI capability and governance change quickly. A benchmark result from 2023 or 2024 may understate current capability, while a successful demonstration or consultation proposal may overstate production maturity or future policy. The paper therefore focuses on architectural patterns that are more stable than model rankings.

Fourth, ``finance'' is not one occupation. Investment banking, accounting, wealth management, trading, risk, insurance, corporate finance, and operations have different data, regulation, client interaction, and error costs.

Finally, this synthesis does not claim that AI will eliminate entry-level finance jobs. It argues that the task bundle is changing and that education should prepare students for a world in which lower-layer information work is increasingly cheap while validation, judgment, governance, and relationships remain scarce.

\section{Conclusion}

The strongest conclusion from recent research is not that finance is becoming fully autonomous. It is that AI is becoming embedded in the workflow between data, software, employees, and customers. Major banks already use generative AI to retrieve internal knowledge, summarize documents, support coding, prepare client follow-up, answer questions, and assist research. Regulators report widespread adoption but very limited full autonomy. Academic research simultaneously shows high task exposure in finance, meaningful productivity gains from human-AI collaboration, uneven reliability across tasks, and emerging pressure on early-career employment in highly exposed occupations.

Connecting these findings changes the career question. The most vulnerable professional is not simply the person who lacks coding skills, nor is the safest professional simply the person with the most finance knowledge. The stronger profile combines domain judgment with the ability to use AI, interrogate it, verify it, and take responsibility for the final result.

For business students, the practical target is therefore not ``learn AI instead of finance.'' It is to learn finance deeply enough to supervise AI, learn data well enough to test it, learn AI well enough to redesign workflows, and develop the communication and judgment needed for decisions that cannot be delegated cleanly.

The next stage of research should measure this transition directly inside finance teams: which tasks disappear, which expand, how apprenticeship changes, and whether AI-literate graduates move faster into higher-value work. Until those data are available, the most defensible interpretation is that AI is compressing the bottom of the Finance Work Stack faster than the top. The future early-career advantage will come from moving upward with it.

\section*{Acknowledgments}
This manuscript is an independent structured evidence synthesis and systems-oriented career framework. No new experimental measurements are claimed. Numerical results and descriptions of existing systems are attributed to the cited researchers, regulators, technology organizations, and financial institutions. Generative AI tools were used to assist literature discovery, source organization, language editing, and LaTeX preparation. The author selected the research question, reviewed the cited evidence, designed the synthesis frameworks, verified the claims included in the manuscript, and accepts responsibility for the complete work.

\bibliographystyle{plainnat}
\bibliography{references}

@article{eisfeldt2025genai,
  title={Generative AI and Finance},
  author={Eisfeldt, Andrea L. and Schubert, Gregor},
  journal={Annual Review of Financial Economics},
  volume={17},
  pages={363--393},
  year={2025},
  doi={10.1146/annurev-financial-112923-020503}
}

@techreport{aldasoro2024intelligent,
  title={Intelligent Financial System: How AI is Transforming Finance},
  author={Aldasoro, Inaki and Gambacorta, Leonardo and Korinek, Anton and Shreeti, Vatsala and Stein, Merlin},
  institution={Bank for International Settlements},
  number={BIS Working Paper No. 1194},
  year={2024},
  url={https://www.bis.org/publ/work1194.htm}
}

@techreport{crisanto2024regulating,
  title={Regulating AI in the Financial Sector: Recent Developments and Main Challenges},
  author={Crisanto, Juan Carlos and Leuterio, Cris Benson and Prenio, Jermy and Yong, Jeffery},
  institution={Financial Stability Institute, Bank for International Settlements},
  number={FSI Insights No. 63},
  year={2024},
  url={https://www.bis.org/fsi/publ/insights63.htm}
}

@techreport{boe2024ai,
  title={Artificial Intelligence in UK Financial Services - 2024},
  author={{Bank of England} and {Financial Conduct Authority}},
  institution={Bank of England and Financial Conduct Authority},
  year={2024},
  url={https://www.bankofengland.co.uk/report/2024/artificial-intelligence-in-uk-financial-services-2024}
}

@article{brynjolfsson2025work,
  title={Generative AI at Work},
  author={Brynjolfsson, Erik and Li, Danielle and Raymond, Lindsey R.},
  journal={The Quarterly Journal of Economics},
  volume={140},
  number={2},
  pages={889--942},
  year={2025},
  note={Earlier version: NBER Working Paper 31161}
}

@techreport{dellacqua2023jagged,
  title={Navigating the Jagged Technological Frontier: Field Experimental Evidence of the Effects of AI on Knowledge Worker Productivity and Quality},
  author={Dell'Acqua, Fabrizio and McFowland III, Edward and Mollick, Ethan R. and Lifshitz-Assaf, Hila and Kellogg, Katherine C. and Rajendran, Saran and Krayer, Lisa and Candelon, Francois and Lakhani, Karim R.},
  institution={Harvard Business School and The Wharton School},
  number={HBS Working Paper 24-013},
  year={2023},
  note={Revised 2026; Organization Science forthcoming},
  url={https://www.hbs.edu/ris/download.aspx?name=24-013.pdf}
}

@techreport{dillon2025shifting,
  title={Shifting Work Patterns with Generative AI},
  author={Dillon, Eleanor and Jaffe, Sonia and Immorlica, Nicole and Stanton, Christopher},
  institution={Microsoft Research},
  year={2025},
  url={https://www.microsoft.com/en-us/research/publication/shifting-work-patterns-with-generative-ai/}
}

@techreport{brynjolfsson2025canaries,
  title={Canaries in the Coal Mine? Six Facts about the Recent Employment Effects of Artificial Intelligence},
  author={Brynjolfsson, Erik and Chandar, Bharat and Chen, Ruyu},
  institution={Stanford Institute for Economic Policy Research},
  year={2025},
  url={https://siepr.stanford.edu/publications/working-paper/canaries-coal-mine-six-facts-about-recent-employment-effects-artificial}
}

@article{bastani2025learning,
  title={Generative AI without Guardrails Can Harm Learning: Evidence from High School Mathematics},
  author={Bastani, Hamsa and Bastani, Osbert and Sungu, Alp and Ge, Haosen and Kabakci, Ozge and Mariman, Rei},
  journal={Proceedings of the National Academy of Sciences},
  volume={122},
  number={26},
  pages={e2422633122},
  year={2025},
  doi={10.1073/pnas.2422633122}
}

@techreport{stanford2026index,
  title={The AI Index Report 2026},
  author={{Stanford Institute for Human-Centered Artificial Intelligence}},
  institution={Stanford University},
  year={2026},
  url={https://hai.stanford.edu/ai-index/2026-ai-index-report}
}

@article{eloundou2023gpts,
  title={GPTs are GPTs: An Early Look at the Labor Market Impact Potential of Large Language Models},
  author={Eloundou, Tyna and Manning, Sam and Mishkin, Pamela and Rock, Daniel},
  journal={arXiv preprint arXiv:2303.10130},
  year={2023}
}

@article{wu2023bloomberggpt,
  title={BloombergGPT: A Large Language Model for Finance},
  author={Wu, Shijie and Irsoy, Ozan and Lu, Steven and Dabravolski, Vadim and Dredze, Mark and Gehrmann, Sebastian and Kambadur, Prabhanjan and Rosenberg, David and Mann, Gideon},
  journal={arXiv preprint arXiv:2303.17564},
  year={2023}
}

@misc{stanfordhelmfinance,
  title={HELM Finance Leaderboard},
  author={{Stanford Center for Research on Foundation Models}},
  year={2025},
  note={Produced in collaboration with Wells Fargo},
  url={https://crfm.stanford.edu/helm/finance/latest/}
}

@techreport{dou2025trading,
  title={AI-Powered Trading, Algorithmic Collusion, and Price Efficiency},
  author={Dou, Winston Wei and Goldstein, Itay and Ji, Yan},
  institution={National Bureau of Economic Research},
  number={Working Paper 34054},
  year={2025},
  note={Revised January 2026},
  url={https://www.nber.org/papers/w34054}
}

@techreport{gambacorta2025relationship,
  title={Artificial Intelligence and Relationship Lending},
  author={Gambacorta, Leonardo and Sabatini, Fabiana and Schiaffi, Stefano},
  institution={Bank for International Settlements},
  number={BIS Working Paper No. 1244},
  year={2025},
  url={https://www.bis.org/publ/work1244.htm}
}

@techreport{bis2025agents,
  title={AI Agents for Cash Management in Payment Systems},
  author={{Bank for International Settlements}},
  institution={Bank for International Settlements},
  number={BIS Working Paper No. 1310},
  year={2025},
  url={https://www.bis.org/publications/working-paper-1310-ai-agents-cash-management-payment-systems}
}

@misc{jpm2025llmsuite,
  title={LLM Suite Named 2025 Innovation of the Year by American Banker},
  author={{JPMorganChase}},
  year={2025},
  month={June},
  url={https://www.jpmorgan.com/technology/news/llmsuite-ab-award}
}

@misc{morgan2023assistant,
  title={Morgan Stanley Wealth Management Announces Key Milestone in Innovation Journey with OpenAI},
  author={{Morgan Stanley}},
  year={2023},
  month={March},
  url={https://www.morganstanley.com/press-releases/key-milestone-in-innovation-journey-with-openai}
}

@misc{morgan2024debrief,
  title={Morgan Stanley Wealth Management Announces AI at Morgan Stanley Debrief},
  author={{Morgan Stanley}},
  year={2024},
  month={June},
  url={https://www.morganstanley.com/press-releases/ai-at-morgan-stanley-debrief-launch}
}

@misc{morgan2024researchgpt,
  title={Morgan Stanley Research Announces AskResearchGPT Powered by OpenAI},
  author={{Morgan Stanley}},
  year={2024},
  month={October},
  url={https://www.morganstanley.com/press-releases/morgan-stanley-research-announces-askresearchgpt}
}

@misc{bofa2025workforce,
  title={AI Adoption by Bank of America's Global Workforce Improves Productivity and Client Service},
  author={{Bank of America}},
  year={2025},
  month={April},
  url={https://newsroom.bankofamerica.com/content/newsroom/press-releases/2025/04/ai-adoption-by-bofa-s-global-workforce-improves-productivity--cl.html}
}

@misc{bofa2025erica,
  title={A Decade of AI Innovation: BofA's Virtual Assistant Erica Surpasses 3 Billion Client Interactions},
  author={{Bank of America}},
  year={2025},
  month={August},
  url={https://newsroom.bankofamerica.com/content/newsroom/press-releases/2025/08/a-decade-of-ai-innovation--bofa-s-virtual-assistant-erica-surpas.html}
}

@misc{goldman2024annual,
  title={Goldman Sachs 2024 Annual Report},
  author={{Goldman Sachs}},
  year={2025},
  note={Describes developer copilot, GS AI Assistant, and AI use cases across businesses},
  url={https://www.goldmansachs.com/investor-relations/financials/current/annual-reports/2024-annual-report}
}

@misc{goldman2025assistant,
  title={Goldman Sachs Launches AI Assistant Firmwide},
  author={{Reuters}},
  year={2025},
  month={June},
  note={Reporting on Goldman Sachs internal memo},
  url={https://www.reuters.com/business/goldman-sachs-launches-ai-assistant-firmwide-memo-shows-2025-06-23/}
}

@misc{bloomberg2025safety,
  title={Bloomberg's Responsible AI Research: Mitigating Risky RAGs and GenAI in Finance},
  author={{Bloomberg}},
  year={2025},
  month={April},
  url={https://www.bloomberg.com/company/stories/bloomberg-responsible-ai-research-mitigating-risky-rags-genai-in-finance/}
}

@misc{wharton2025adoption,
  title={Accountable Acceleration: Gen AI Fast-Tracks Into the Enterprise - 2025 AI Adoption Report},
  author={Korst, Jeremy and Puntoni, Stefano and Tambe, Prasanna},
  year={2025},
  institution={Wharton Human-AI Research and GBK Collective},
  url={https://knowledge.wharton.upenn.edu/special-report/2025-ai-adoption-report/}
}

@misc{hbs2025tutor,
  title={What Happens When Every Student Has an AI Tutor?},
  author={{Harvard Business Publishing Education}},
  year={2025},
  month={August},
  note={Describes analysis of more than 20,000 prompts from 930 HBS accounting students},
  url={https://he.hbsp.harvard.edu/2025-08-05-the-faculty-lounge.html}
}

@article{hirabayashi2024harvard,
  title={Harvard Undergraduate Survey on Generative AI},
  author={Hirabayashi, Shikoh and Jain, Rishab and Jurkovic, Nikola and Wu, Gabriel},
  journal={arXiv preprint arXiv:2406.00833},
  year={2024}
}

@misc{booth2025education,
  title={Bringing AI into Business Education: The Future of Finance},
  author={{Chicago Booth Center for Applied Artificial Intelligence}},
  year={2025},
  month={September},
  url={https://www.chicagobooth.edu/research/center-for-applied-artificial-intelligence/stories/2025/bringing-ai-into-business-education}
}

@misc{boothwithdrawal,
  title={Ongoing Research Projects: Notice on Withdrawn LLM Financial Statement Analysis Working Papers},
  author={Nikolaev, Valeri V.},
  year={2026},
  institution={University of Chicago Booth School of Business},
  url={https://faculty.chicagobooth.edu/valeri-nikolaev/ongoing-research-projects}
}

@misc{mastercard2025fraud,
  title={The Right Side of AI: Shaping the Future of Payment Fraud Prevention},
  author={{Mastercard}},
  year={2025},
  url={https://www.mastercard.com/content/dam/mccom/shared/news-and-trends/insights/2025/one-the-right-side-of-ai/the-right-side-of-ai-shaping-the-future-of-payment-fraud-prevention.pdf}
}

@techreport{fsb2026sound,
  title={Sound Practices for Responsible Adoption of Artificial Intelligence (AI): Consultation Report},
  author={{Financial Stability Board}},
  institution={Financial Stability Board},
  year={2026},
  month={June},
  note={Consultation report proposing 12 practices covering organization-wide governance and AI lifecycle, cyber, ICT, and third-party risk},
  url={https://www.fsb.org/2026/06/sound-practices-for-responsible-adoption-of-artificial-intelligence-ai-consultation-report/}
}

@misc{fsb2026frontier,
  title={FSB Chair's Letter to G20 Finance Ministers and Central Bank Governors: August 2026},
  author={{Financial Stability Board}},
  year={2026},
  month={August},
  note={Warns that frontier AI may materially alter the speed, scale, and economics of cyber risk},
  url={https://www.fsb.org/2026/08/fsb-chairs-letter-to-g20-finance-ministers-and-central-bank-governors-august-2026/}
}

\end{document}